\documentstyle[aps,prl,psfrag,epsf]{revtex}

\newcommand{\bea}{\begin{eqnarray}}
\newcommand{\eea}{\end{eqnarray}}

\newcommand{\ak}{a^{\dagger}}
\newcommand{\ket}[1]{| #1 \rangle}
\newcommand{\bra}[1]{\langle #1 |}

\begin{document}
\def\overlay#1#2{\setbox0=\hbox{#1}\setbox1=\hbox to \wd0{\hss #2\hss}#1
-2\wd0\copy1}
\twocolumn[\hsize\textwidth\columnwidth\hsize\csname@twocolumnfalse\endcsname

\title{Quantum systems coupled to a
  structured reservoir with multiple excitations.}
\author{Georgios M. Nikolopoulos$^{1,2}$, S\o ren Bay$^1$ 
and P. Lambropoulos$^{1,2}$}

\address{
  1. Max-Planck-Institut f\"ur Quantenoptik,
  Hans-Kopfermann-Str.\ 1, 85748 Garching, Germany\\
  2. Institute of Electronic Structure {\rm \&} Laser, FORTH,
  P.O. Box 1527, Heraklion 71110, Crete, Greece;\\
  and Department of Physics, University of Crete, Crete, Greece}
\date{\today}
\maketitle
\begin{abstract}

We present a method for dealing with quantum systems coupled to a 
structured reservoir with any density of modes and  
with more than one excitation. We apply the method to a two-level atom
coupled to the edge of a photonic band gap and a defect mode. Results
pertaining to this system, provide the solution to the problem of
two photons in the reservoir and possible generalization is discussed.

\end{abstract}
\vskip2pc]

\tighten

The problem of the interaction of small systems with structured
reservoirs is of central importance to a number of areas including
nanostructures in semiconductors, atom lasers \cite{savage}, 
 aspects of molecular dynamics \cite{domcke} and atoms
embedded in photonic band gap materials. 
A fundamental difficulty in the theoretical formulation of such problems
stems from the invalidation of the Born-Markov approximations,
 essential in obtaining a master equation, which is the standard
vehicle in the presence of smooth reservoirs. Models approximating
some of the features of such reservoirs such as superpositions of
Lorentzians can be useful, as the introduction of ``pseudo-modes'' can
lead to a Markovian master equation for a system slightly enlarged
through the introduction of the pseudo-modes
\cite{imamoglu,imamogluqso97,garraway,baysuper}. 
Alternatively, one may
introduce decorrelation approximations in the Heisenberg equations of
motion for
the operators of interest. But in any case, one can not be confident
of the validity and degree of accuracy of the approximations. 
For a general density of states that can not be modelled by a
superposition of Lorentzians, the dynamics can only be obtained if
there is at most one photon in the structured reservoir \cite{baylambdaprl}.
At this point, there is no generally established 
approach that can provide a description of
the dynamics for a general density of states and multiple excitations
in the structured continuum.\\
It is the purpose of this paper to present such an approach with 
illustrative applications. The basic idea relies on the
discretization of the continuum which is thus replaced in the
formulation by a finite (but large) number of discrete modes. 
Their couplings and frequencies are chosen so as to model the
effect of the structured continuum to the desired accuracy. The
judicious choice of this parametrization is of critical importance to
the success of this idea. Given the discretization, the system
``atom+discretized continuum'' can be handled through differential
equations governing the evolution of the amplitudes entering the
Schr\"odinger equation. These differential equations are then solved
numerically. The discretization of continua in other contexts
\cite{twoelectron} is an
established but always dangerous approach requiring much care as it
can lead to unphysical artifacts.

To introduce and 
demonstrate the method, we consider a two-level atom (TLA) coupled
near-resonantly to the edge of a photonic band gap (PBG)
\cite{yablo87,johnprl87}.
The photonic band gap material has
a strongly modified dispersion relation and employing the isotropic
dispersion relation introduced by John and Wang \cite{johnprb}, the 
corresponding density of states reads
\begin{eqnarray}
\rho(\omega)=\frac{k}{\sqrt{\omega-\omega_e}}
\Theta(\omega-\omega_e) \label{dos}
\end{eqnarray}
where $k$ is a material specific constant, $\omega_e$ is the band edge
frequency and $\Theta(x)$ is the Heaviside step-function.
Clearly, the density of modes diverges at the edge
frequency which invalidates the standard Born-Markov approximations
normally employed when dealing with a smooth reservoir. As a consequence
the reservoir can not be eliminated.
\\
The idea is to replace the density of modes in Eq. (\ref{dos}) near the 
atomic transition (which for our purposes will be in the vicinity of the edge
frequency), by a collection of discrete harmonic 
oscillators, while the rest of the mode-density can be treated perturbatively 
since it is far from resonance.
The frequencies and the couplings of the discrete modes are chosen such that 
the discrete oscillators best model the structured continuum near the edge 
frequency. 
To this end, we write
Eq. (\ref{dos}) in a differential form
\bea
\Delta N=\rho(\omega)\Delta \omega \label{deltan}
\eea
For $\Delta N=1$ and introducing a discrete index, we find
$\Delta \omega_i=1/\rho(\omega_i)$
and thus
\bea
\omega_{i+1}=\omega_i+\Delta\omega_i=\omega_i+1/\rho(\omega_i) \label{disc1}
\eea
and $\omega_1=\omega_e+\delta$ where $\delta$ is chosen sufficiently small
($\delta\approx 10^{-2}C^{2/3}$).
The coupling $g_r$ to the discrete modes is found by integration of
Eq. (\ref{deltan}) 
\bea
\sum g_r^2 \Delta N \approx 
\int_{\omega_e}^{\omega_u} d\omega |\kappa_\omega|^2\rho(\omega)
\eea
where $\omega_u$ is the upper limit of the discretized part of the density of 
states, $\kappa_\omega$ is
the coupling between the continuum mode with frequency 
$\omega$ and the atom and 
$|\kappa_\omega|^2\rho(\omega)=\frac{C}{\pi}
\frac{1}{\sqrt{\omega-\omega_e}}$,
where $C$ is the effective coupling of the atom to the PBG
structure. We thus find
\bea
g_r\approx\sqrt{ \frac{2C}{N\pi}\sqrt{\omega_u-\omega_e}} \label{coupl}
\eea 
where $N$ is the number of discrete modes.
An alternative form of Eq. ({\ref{disc1}}) is
\bea
\omega_{i+1}=\omega_{i-1}+\Delta\omega_i=\omega_{i-1}+2/\rho(\omega_i) 
\label{disc2}
\eea
which is actually the form that we have used in our implementation.\\
We consider a two-level atom with ground $(\ket{g})$ and excited 
$(\ket{e})$ states whose energy difference is $\hbar \omega_o$.
The atom is coupled to the structured reservoir and a defect mode centered at 
the frequency $\omega_d$ inside the gap. 
The Hamiltonian for this system in an interaction
picture rotating at the band edge frequency $\omega_e$ ($\hbar=1$) and in the
rotating wave approximation reads
\begin{eqnarray}
H&=&\Delta_o\sigma_{ee}+\Delta_d\ak_d a_d+\sum_\lambda \Delta_\lambda
\ak_\lambda a_\lambda+g_d (a_d\sigma^+ +\ak_d \sigma^- )\nonumber \\
&&+\sum_\lambda g_\lambda (a_\lambda\sigma^+ +\ak_\lambda \sigma^- )
\end{eqnarray}
where $\Delta_o=\omega_o-\omega_e$, $\Delta_d=\omega_d-\omega_e$,
$\Delta_\lambda=\omega_\lambda-\omega_e$;
$\sigma^+=\ket{e}\bra{g}$ and $\sigma^-=\ket{g}\bra{e}$ 
are the atomic raising and lowering operators and 
$\sigma_{ee}=\sigma^+\sigma^-$. The field operators $(a_d,\ak_d)$ and
$(a_\lambda,\ak_\lambda)$ correspond to the defect mode and PBG reservoir, 
respectively, which are coupled to the atom via the respective coupling
constants $g_d$ and $g_\lambda$.
\\ 
In order to demonstrate the validity of this method, we first present
the results for spontaneous decay i.e. the atom is initially excited,
and we neglect the defect mode i.e. $g_d=0$.  
Replacing the density of modes of Eq. (\ref{dos}) for $\omega<\omega_u$, 
by a collection of discrete modes, the wavefunction for 
the full system reads
\begin{eqnarray}
\ket{\psi}=a_0\ket{e,0}+\sum_{j} b_{j}\ket{g,1_{j}}+\sum_{\lambda} b_{\lambda}\ket{g,1_{\lambda}}
\end{eqnarray}
where the amplitudes $b_j$ correspond to the discrete modes, while the
$b_{\lambda}$ correspond to the modes with frequency 
$\omega_{\lambda}>\omega_u$, which are treated pertubatively 
i.e. they are eliminated adiabatically.\\
The time evolution of the amplitudes are governed by the Schr\"odinger
equation from which we obtain
\begin{eqnarray}
\dot{a}_0&=&\frac{1}{i}\Delta_o a_0+\frac{1}{i}\sum_{j=1}^N g_j
b_j+\frac{1}{i}\sum_{\lambda} g_{\lambda} b_{\lambda} \label{sda0}
\\
\dot{b}_j&=&\frac{1}{i}\Delta_j b_j+\frac{1}{i}g_j a_0 \label{sdbj}
\\
\dot{b}_{\lambda}&=&\frac{1}{i}\Delta_{\lambda} b_{\lambda}+
\frac{1}{i}g_{\lambda} a_0 \label{sdblambda}
\end{eqnarray}

\begin{figure}
  \begin{center}
    \leavevmode
    \epsfxsize8.5cm
    \epsfbox{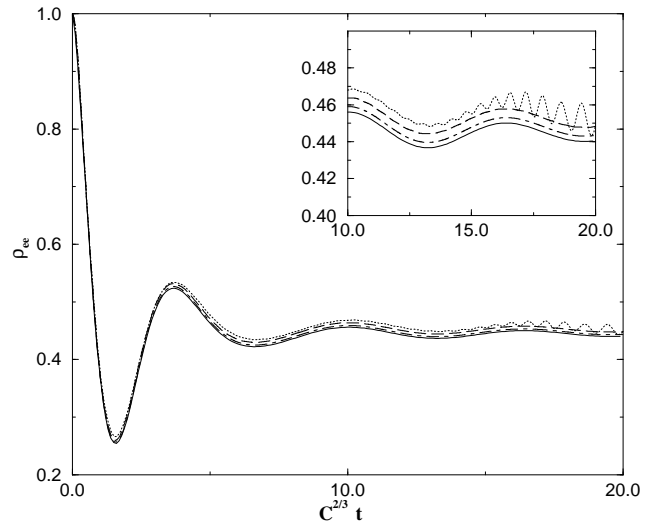}
  \end{center}
\caption{The population in the excited state as function of time. The
    solid line is the exact solution. The dotted line is for $N=50$.
    The long-dashed line is for $N=150$ and the
    dotted-dashed line is for $N=500$. The insert shows a
    close-up of the long-time behavior. Parameters:
    $\Delta_o=0$ and $g_d=0$ }
\label{spontdec.fig}
\end{figure}

\begin{figure}
  \begin{center}
    \leavevmode
    \epsfxsize9.5cm
    \epsfbox{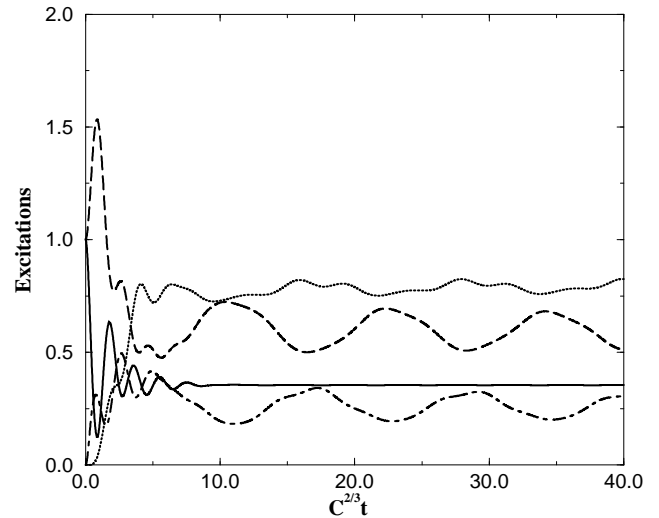}
  \end{center}
\caption{The evolution of the system is plotted as function of
    time. The solid line is the population in the upper atomic
    state. The long-dashed line is the mean photon number in the
    defect mode, the dot-dashed line is the population in the
    one-photon sector of the reservoir Hilbert space and the dotted
    curve is the population in the two-photon sector of the reservoir
    Hilbert space. Parameters: $N=150$, $g_d=C^{2/3},
    \Delta_o=\Delta_d=-0.1C^{2/3}$}
\label{twop.fig}
\end{figure}

Formal integration of Eq. (\ref{sdblambda}) gives
\begin{eqnarray}
b_{\lambda}(t)-b_{\lambda}(t_0)e^{\Delta_{\lambda} (t-t_0)/i}
=\frac{g_{\lambda}}{i}\int_{t_0}^{t}dt' a_0(t') e^{\Delta_{\lambda}(t-t')/i}
\end{eqnarray}
Since these modes are strongly off-resonant i.e. 
$\Delta_{\lambda}\gg g_{\lambda}$  
and for short times $a_0(t')$ remains
almost constant, $a_0(t')$ can be replaced by $a_0(t)$. The remaining
integral over the exponential is easily performed with the result
\begin{eqnarray}
b_{\lambda}(t)\simeq \frac{g_{\lambda}}{i^2\Delta_{\lambda}}a_0(t).\label{elim}
\end{eqnarray}
Substituting Eq. (\ref{elim}) into Eq. (\ref{sda0}) we have
\begin{eqnarray}
\dot{a}_0&=&\frac{1}{i}\Delta_o a_0+\frac{1}{i}\sum_{j=1}^N g_j b_j-
\sum_{\lambda}\frac{g_{\lambda}^2}{i\Delta_{\lambda}}a_0 \label{sda02}
\\
\dot{b}_j&=&\frac{1}{i}\Delta_j b_j+\frac{1}{i}g_j a_0 \label{sdbj2}
\end{eqnarray}
Converting the mode sum over ${\lambda}$ into an integral from 
$\omega=\omega_u$ to infinity and using Eq. (\ref{dos}), we obtain
\begin{eqnarray}
\dot{a}_0&=&\frac{1}{i}\left(\Delta_o-
\frac{g_{j}^2 N}{\omega_u-\omega_e}\right)a_0+
\frac{1}{i}\sum_{j=1}^N g_j b_j \label{sda02}
\\
\dot{b}_j&=&\frac{1}{i}\Delta_j b_j+\frac{1}{i}g_j a_0 \label{sdbj2}
\end{eqnarray}
where for all discretized modes $g_j=g_r$ as given in Eq. (\ref{coupl}).
The effect of the smoothly varying part of the density of modes is thus to
add a vacuum shift term to the equation of motion for the upper state
amplitude which effectively shifts the level down in energy and thus
towards the band gap where it is protected from decay. 
This approximation leads to a significantly reduced number of differential
equations and the remaining amplitudes are distributed
over a much narrower frequency interval. Beyond that, the approximation 
also provides a surprising insight into the physical process, 
as discussed above.
\\
To ensure satisfactory numerical agreement with the known exact 
solution \cite{johnsd,baytaipra} for this test problem, we find that 
we need at least 150 modes. 
In Fig. \ref{spontdec.fig}, we present the results obtained by
propagation of Eqs. (\ref{sda02}) and (\ref{sdbj2}). The dotted line is
for a calculation with 50 discrete modes, the long-dashed line is
for 150 discrete modes, and the dashed-dotted line is for 500 modes.  
For comparison, we also plot the exact known
solution \cite{johnsd,kurizkisd,baytaipra} (solid line) which shows very good 
agreement
with the calculation involving 150 modes (estimated error $2\%$). 
The curve corresponding to
500 modes is indistinguishable from the exact solution. The
calculation involving 50 modes, exhibits revivals for longer
times. These are a consequence of the discretization; one of the
dangerous artifacts that one must be cognizant of.  Increasing the
number of discrete modes, the revivals appear for later and later
times. The number of modes in our calculations thus determines the time
scale on which the propagation is free of artificial oscillations, while 
$\omega_u$ determines the proximity of the envelope to the correct result. 
This implies considerable 
flexibility in the method; in the sense that the size of the calculation
can be tailored to the time scale, over which the behavior of the system
is sought and the desired accuracy.
\\
Having demonstrated the validity of the method, 
we address now an open problem. Adding to the system described above a
defect mode near-resonant with the atom, this defect mode 
acts as a photon source that can pump the atom. With
one photon in the defect mode and the atom excited at $t=0$, we have the
possibility of two photons in the reservoir, a problem not amenable to 
techniques employed so far. The wavefunction for the system can be written
\begin{eqnarray}
\ket{\Psi(t)}=a_0\ket{e,1_d,0}+b_0\ket{g,2_d,0}
+\sum_j b_j\ket{g,1_d,1_j}\nonumber\\
+\sum_j a_j\ket{e,0,1_j}
+\sum_{j,k} b_{j,k}\ket{g,0,1_j,1_k}
\end{eqnarray}
where the states involved are product states and, for instance,
$\ket{g,1_d,1_j}=\ket{g}\ket{1_d}\ket{1_j}$ where $\ket{1_d}$ is
the one-photon state of the defect mode and $\ket{1_j}$ is a
one-photon state of the reservoir.
The amplitudes obey the Schr\"odinger equation, and through the
perturbative elimination of off-resonant modes as described above, we
find
\begin{eqnarray}
\dot{a}_0&=& \frac{1}{i}\left(\Delta_o+\Delta_d-
\frac{g_{j}^2 N}{\omega_u-\omega_e}\right)a_0 \nonumber\\
&&+\frac{1}{i}\sqrt{2}g_db_0+\frac{1}{i}\sum_{j=1}^N g_j b_j 
\\
\dot{b}_0&=&\frac{2}{i}\Delta_d b_0+\frac{1}{i}\sqrt{2}g_d a_0 
\\
\dot{a}_j&=&\frac{1}{i}\left(\Delta_o+\Delta_j-
\frac{g_{k}^2 N}{\omega_u-\omega_e}\right)a_j+\frac{1}{i}g_d b_j \nonumber\\
&&+\frac{1}{i}\sum_{\stackrel{k=1}{(k\neq j)}}^{N} g_k b_{jk}
+\frac{1}{i}\sqrt{2}g_j b_{jj}
\\
\dot{b}_j&=& \frac{1}{i}(\Delta_j+\Delta_d)b_j+\frac{1}{i}g_j a_0+
\frac{1}{i}g_da_j 
\\
\dot{b}_{jk}&=&\frac{1}{i}(\Delta_k+\Delta_j)b_{jk}+\frac{1}{i}g_ka_j
+\frac{1}{i}g_ja_k \\
\dot{b}_{jj}&=&\frac{2}{i}\Delta_jb_{jj}+\frac{1}{i}\sqrt{2} g_ja_j
\end{eqnarray}
where $j,k$ are mode indices and for all discretized modes $g_j=g_k=g_r$. 
For the purposes of this example, the frequency 
$\omega_d$ of the defect mode is inside the gap as determined by the value
of $\Delta_d$ and the atomic transition on resonance with $\omega_d$, i.e. 
$\Delta_o=\Delta_d$.\\
This set of equations is solved numerically with the results
presented in Fig. \ref{twop.fig}. We plot the
atomic inversion (solid line), the mean-photon number in the defect
mode (long-dashed line) and the populations in the one-photon sector
(dot-dashed line) and two-photon sector (dotted line) of the reservoir
Hilbert space, respectively, as functions of time. 
From the figure, we find that there is an exchange
of energy (oscillation) 
between the defect mode and the one-photon sector of the
reservoir. This oscillation must involve the atom, since
the defect mode is not directly coupled to the reservoir, but is not
reflected in the atomic inversion. As is evident in
Fig. \ref{twop.fig}, although photons are exchanged between the defect
mode and the reservoir through the atom, after some initial time, the
atomic population remains practically constant; a rather surprising effect. 
The results presented in Fig. 2, have of course been tested for convergence
in terms of number of modes, $\omega_u$, etc.\\ 
As the defect mode is pushed further into the gap, we find that the 
oscillations of the atomic population begin to extend to increasingly longer
times. On the contrary, a change in the magnitude of $g_d$ in relation to
$C^{2/3}$ does not seem to affect the atomic oscillations for longer times,
but it does affect the relative oscillations of the excitations in the defect 
mode and the reservoir, as we will discuss in detail elsewhere. 
\\ \\
In conclusion, we have developed an approach that is capable of
providing solutions to a class of problems which can only be treated
approximately through other techniques. It is applicable to small
systems coupled to a density of modes of any form and has allowed us
to solve problems involving multiple excitations in the continuum. In
addition to the implementation outlined here, we have explored various
other forms of discretizations, as well as other densities of modes,
with good agreement with other exact results in those cases that are
available. This demonstrates the generality and versatility of the
approach, which could be readily employed in other contexts such as, for 
example, waveguides where the density of modes is also singular.
\cite{klep,lew} \\
We demonstrated, in addition, that
off-resonant continuum modes can be eliminated perturbatively.
 The effect of this approximation is to reduce the number of
differential equations to be propagated, leading thus to a drastic
enhancement of the computation speed, essentially without compromise in
accuracy. 
The number of equations to propagate scales roughly
as $N^{p}$, where $N$ is the number of discrete modes
and $p$ the number of excitations.
The ultimate limitation of the method is determined by computer memory
as demanded by each problem. In particular, the study of multiple
excitations will probably for the time being be limited to 4 or 5.
For the purpose of
presenting the method, we limited our discussion here to two
photons. Results from work for more photons will be presented
elsewhere, but we do mention here that we have also obtained fully converged
results for three photons in the reservoir.
\\
Providing solutions for the dynamics of the system is one aspect of
this approach. Perhaps an equally important aspect is the insight
gained by the possibility to combine the perturbative treatment with
the non-perturbative treatment of the rest. As discussed above, this
sheds light on the physical effect of the modes around the band edge
as compared to the smooth distant part of the density of modes.
In addition, the possibility to monitor the dynamics of additional photons
in the structured reservoir may prove very valuable when considering
the validity of approximations necessary in other schemes.
\\
It should be mentioned in closing that a recently proposed formal
approach to similar non-Markovian problems, based on the quantum state 
diffusion formalism, has been presented by
Diosi, Strunz and Gisin \cite{diosi}. At this point we
are aware of the application of the method to a relatively tractable
problem involving a standard cavity reservoir. Its potential, however,
does not seem at first sight to be limited and it will be interesting 
to see and explore
its applications to problems involving more complicated
densities of modes.

\bibliographystyle{prsty}

\end{document}